\documentclass[12pt]{article}
\usepackage{amsmath,amsfonts}
\makeatletter \@addtoreset{equation}{section}

\usepackage{cite}
\usepackage{bm}
\usepackage{dcolumn}
\makeatother
\def\one{{\hbox{ 1\kern-.8mm l}}}
\newcommand{\Dslash}{\not{\hbox{\kern-4pt $D$}}}
\newcommand{\pdslash}{\not{\hbox{\kern-2pt $\partial$}}}
\newcommand{\be}{\begin{equation}}
\newcommand{\bea}{\begin{eqnarray}}
\newcommand{\eea}{\end{eqnarray}}
\newcommand{\ba}{\begin{array}}
\newcommand{\ea}{\end{array}}
\newcommand{\ee}{\end{equation}}

\begin{document}

\begin{titlepage}
\vspace*{1mm}%
\hfill%
\vspace*{15mm}%
\begin{center}

{{\Large {\bf On Holographic Realization of Logarithmic GCA }}}

\vspace*{15mm} \vspace*{1mm} {Ali Hosseiny $^{a}$ and Ali
Naseh$^{a,b}$}

 \vspace*{1cm}

{\it ${}^a$ Department of Physics, Sharif University of Technology \\
P.O. Box 11365-9161, Tehran, Iran }

\vspace*{.4cm}

{\it ${}^b$ School of physics, Institute for Research in Fundamental Sciences (IPM)\\
P.O. Box 19395-5531, Tehran, Iran \\}

\vspace*{.4cm}

email: alihosseiny@physics.sharif.edu, and naseh@ipm.ir

\vspace*{2cm}
\end{center}

\begin{abstract}

We study 2-dimensional Logarithmic Galilean Conformal Algebra (LGCA)
by making use of a contraction of Topologically Massive Gravity at
critical point. We observe that using a naive contraction at the
critical point fails to give a well defined theory, though
contracting the theory while we are approaching the critical point
leads to a well behaved expression for two point functions of the
energy-momentum tensors of LGCA.

\end{abstract}

\end{titlepage}



\section{Introduction}

Beside the Schrodinger algebra
\cite{{Niederer:1972zz},{Hagen:1972pd},{Henkel:1993sg}} which is
the most celebrated non-relativistic conformal algebra, another
non-relativistic algebra named GCA \cite{Havas:1978} has recently
received a lot of attention \cite{{Alishahiha:2009np},
{Bagchi:2009my}}. GCA is the only known
non-relativistic conformal symmetry which scales space and time
isotropically

\bea
x\rightarrow \lambda x, \;\;\;\;\;\;\;\;\;\;\;\;\;\;\;\;\; t \rightarrow \lambda t.
\eea

Usually non-relativistic conformal symmetries scale space and time
anisotropicaly

\bea
x\rightarrow \lambda x, \;\;\;\;\;\;\;\;\;\;\;\;\;\;\;\;\; t \rightarrow \lambda^{\theta} t.
\eea

$\theta$ is called the anisotropy index. Looking for the most general
non-relativistic conformal symmetries in $d\neq 2$ one ends up with
the class of l-Galilei algebras\cite{Henkel:1997} \footnote{ For two
spatial dimensions a larger class has been claimed for the sake of
respecting infinite dimensional symmetries of $CFT_{2}$ in $2$
spatial dimensions \cite{Hosseiny:2009jj}}. Each element of the
class is identified with a half-integer number $l$ which is the
inverse of $\theta$. Being isotropic in scaling space and time is
not GCA's only special feature. GCA is as well unique by being
obtained from contracting relativistic conformal symmetry in spatial
dimensions.

\bea\label{con.def} x\rightarrow
\frac{x}{c},\;\;\;\;\;\;\;\;\;\;\;\;\;\;\;\;\; t \rightarrow
t.\;\;\;\;\;\;\;\;\;\;\;\;\;\;\;\;\;\;c\rightarrow \infty \eea

From a physical point of view it means that we investigate the
behavior of system for low speeds or maybe low energies. GCA can be
extended to an infinite dimensional algebra which is named full GCA.
In $1+1$- dimensions full GCA can be obtained from contraction of
conformal symmetry which is two Virasoros \cite{Hosseiny:2009jj}.
This special feature of full GCA in $1+1$-dimensions helps to know
its quantum behavior from contracting $CFT_{2}$
\cite{{Bagchi:2009pe},{Hosseiny:2010sj}}.\\ Beside regular conformal
field theories which are unitary, there is another class of
conformal models which are not unitary and named logarithmic
conformal field theorise (LCFT). LCFT's arise when the action of
scaling operator on scaling fields are not diagonal but rather
Jordan \cite{Gurarie:1993xq}

\bea
L_{0}\phi\;=\;h\phi,\;\;\;\;\;\;\;L_{0}\psi\;=\;h\psi+\kappa\phi.
\eea

Although being non-unitary, LCFT's have a large variety of
applications in field theories and statistical physics\footnote{ For
example see
\cite{Flohr:2001zs,MoghimiAraghi:2000qn,Mathieu:2007pe,Henkel:2010}
and references therein.}. As regular $CFT_{2}$, LCFT's can
be contracted and yield logarithmic GCA (LGCA) \cite{Hosseiny:2010sj}.\\
One of the suggested realizations of $CFT_{2}$ in the context of
holography is Topologically Massive Gravity (TMG)
\cite{{Deser:1981wh},{Deser:1982vy}} away from the critical point.
Recently this correspondence has been looked from the contraction
point of view to suggest duality between contracted TMG and GCA
\cite{Bagchi:2010mw}. It is shown that TMG at the critical point
might be dual to LCFT \cite{{Skenderis:2009nt},{Grumiller:2009mw}}. In this paper we
search the possibility of correspondence between LGCA on the
boundary and contracted TMG at the critical point in the bulk. In
section 2 we review GCA and its contraction from $CFT_{2}$ and as
well its logarithmic representation. In section $3$ we review
Topologically Massive Gravity in critical point. In section $4$ we
utilize contraction approach to observe holographic realization of
LGCA.


\section{GCA and Contraction}

Up to our knowledge the oldest reference to GCA is
\cite{Havas:1978}. It as well were investigate later and called
$\mathfrak{alt}$ in \cite{Henkel:2003pu}. Its exotic central charge
has been of interest
\cite{Duval:2000xr,Jackiw:1990ka,Lukierski:2005xy}. Later it got of
interest in the context of holography
\cite{Alishahiha:2009np,Bagchi:2009pe}. Full GCA which is infinite
extension of GCA is represented by the following operators \cite{Henkel:2002vd}

\bea\begin{split}
&T^{n}\;=\; -(n+1)t^{n}x_{i}\partial_{i}-t^{n+1}\partial_{t},&\cr&
M_{i}^{n}\;=\; t^{n+1}\partial_{i},&\cr&
J_{ij}^{n}\;=\; -t^{n}(x_{i}\partial_{j}-x_{j}\partial_{i}).
\end{split}\eea

This infinite algebra commutes as

\bea\begin{split}
&[T^{m},T^{n}]\;=\;(m-n) T^{m+n},\;\;\;\;\;\;\;[T^{m},J_{a}^{n}]\;=\;-n J_{a}^{m+n},&\cr&
[J_{a}^{m},J_{b}^{n}]\;=\; f_{abc} J_{c}^{m+n},\;\;\;\;\;\;\;\;\;\;\;\;\;\;\;[T^{m},M_{i}^{n}]\;=\;(m-n)M_{i}^{m+n},&\cr&
[M_{i}^{m},M_{j}^{n}]\;=\;0,\;\;\;\;\;\;\;\;\;\;\;\;\;\;\;\;\;\;\;\;\;\;\;\;
[M_{i}^{m},J_{jk}^{n}]\;=\;(M_{j}^{m+n}\delta _{ik}-M_{k}^{m+n}\delta_{ij}).
\end{split}\eea

It can be observed that full GCA in $1+1$-dimensions can be obtained
directly from contracting conformal symmetry in $2$-dimensions
\cite{Hosseiny:2009jj}. In fact in two dimensions

\bea
z\;=\;x+t,\;\;\;\;\;\;\;\;\;\;\;\;\;\;\;\;\;\overline{z}\;=\;x-t.
\eea

conformal symmetry is two Virasoro algebra

\bea
L_{n}\;=\;-z^{n+1}\partial_{z},\;\;\;\;\;\;\;\;\;\;\overline{L}_{n}\;=\;-\overline{z}^{n+1}\partial_{\overline{z}}.
\eea

which commute as

\bea\label{V.a}\begin{split} &
[L_{m},L_{n}]\;=\;(m-n)L_{m+n}+\frac{c_{L}}{12}\;m(m^{2}-1)\delta_{m+n,0},&\cr&
[\overline{L}_{m},\overline{L}_{n}]\;=\;(m-n)\overline{L}_{m+n}+\frac{c_{R}}{12}\;m(m^{2}-1)\delta_{m+n,0}.
\end{split}\eea

In the contraction limit we observe

\bea\begin{split}
&L_{n}\;=\;-(t+\frac{x}{c})^{n+1}(\partial_{t}-c\partial_{x})
\;=\;-t^{n+1}(-c\partial_{x}+\partial_{t}+(n+1)\frac{x}{t}\partial_{t}+O(\frac{1}{c})),&\cr&
\overline{L}_{n}\;=\;-(-t+\frac{x}{c})^{n+1}(-\partial_{t}-c\partial_{x})
\;=\;-t^{n+1}(c\partial_{x}+\partial_{t}+(n+1)\frac{x}{t}\partial_{t}+O(\frac{1}{c})).
\end{split}\eea

Redefining operators

\bea
T_{n}\;=\;L_{n}+\overline{L}_{n},\;\;\;\;\;\;\;\;\;\;\;\;M_{n}\;=\;\frac{L_{n}-\overline{L}_{n}}{c}.
\eea

we end up with full GCA. Since we know the field theory of conformal
symmetry in $2$ dimensions via quantizing symmetry in (\ref{V.a}) we
can follow contraction and obtain quantized GCA in $1+1$-dimensions
\cite{Bagchi:2009pe}. Considering the states of\;$CFT_{2}$\;which
are characterized by their holomorphic and antiholomorphic scaling
weights $|h,\overline{h}\rangle$ in which

\bea
L_{0}|h,\overline{h}\rangle\;=\;h|h,\overline{h}\rangle,\;\;\;\;\;\;\;\;\;\;\;
\overline{L}_{0}|h,\overline{h}\rangle\;=\;\overline{h}|h,\overline{h}\rangle.
\eea

one can obtain states of $GCA_{2}$

\bea\begin{split}
&T_{0}|h,\overline{h}\rangle=(L_{0}+\overline{L}_{0})|h,\overline{h}\rangle
\;=\;(h+\overline{h})|h,\overline{h}\rangle,&\cr&
M_{0}|h,\overline{h}\rangle\;=\;\frac{L_{0}-\overline{L}_{0}}{c}|h,\overline{h}\rangle\
\;=\;\frac{h-\overline{h}}{c}|h,\overline{h}\rangle.
\end{split}\eea

As we observe, scaling states of $CFT_{2}$ are scaling states of GCA
too. They are identified by their scaling weight and rapidity

\bea T_{0}|\mathcal{H} ,\xi\rangle\;=\;\mathcal{H}|\mathcal{H}
,\xi\rangle,\;\;\;\;\;\;\;\;M_{0}|\mathcal{H}
,\xi\rangle\;=\;\xi|\mathcal{H},\xi\rangle. \eea

in which

\bea
\mathcal{H}\;=\;h+\overline{h},\;\;\;\;\;\;\;\;\;\;\;\xi\;=\;\frac{1}{c}(h-\overline{h}).
\eea

We have then

\bea
h\;=\;\frac{1}{2}(\mathcal{H}+c\xi),\;\;\;\;\;\;\;\;\;\;\;\;\overline{h}\;=\;\frac{1}{2}(\mathcal{H}-c\xi).
\eea

Looking at commutation relations of $CFT_{2}$ generators (\ref{V.a})
one can find new GCA central charges

\bea C_{1}\;=\; c_{R}+c_{L},\;\;\;\;\;\;\;\;\;\;\;C_{2}\;=\;
\frac{c_{R}-c_{L}}{c}, \eea

in which

\bea\begin{split}
&[T_{m},T_{n}]\;=\;(m-n)T_{m+n}+\frac{C_{1}}{12}\;m(m^{2}-1)\delta_{m+n,0},&\cr&
[T_{m},M_{n}]\;=\;(m-n)M_{m+n}+\frac{C_{2}}{12}\;m(m^{2}-1)\delta_{m+n,0},&\cr&
[M_{m},M_{n}]\;=\;0.
\end{split}\eea

As it can be observed, to have finite and nonzero $\mathcal{H},\xi$,
one need to have large $h,\overline{h}$ and $c_{R},c_{L}$ in
opposite signs.\\
As regular CFT the logarithmic CFT can be contracted and yield LGCA
. If we consider LCFT scaling states

\bea\begin{split} &L_{0}|h,0\rangle\;=\;h|h,0\rangle,&\cr&
L_{0}|h,1\rangle\;=\;h|h,1\rangle+\kappa|h,0\rangle.
\end{split}\eea

and set $\kappa=1$ then under contraction procedure we find
\cite{Hosseiny:2010sj}

\bea\begin{split}
&T_{0}|h,\overline{h},0\rangle\;=\;(h+\overline{h})|h,\overline{h},0\rangle
\;=\;\mathcal{H}|\mathcal{H},\xi,0\rangle,&\cr&
M_{0}|h,\overline{h},0\rangle\;=\;\frac{h-\overline{h}}{c}|h,\overline{h},0\rangle
\;=\;\xi|\mathcal{H},\xi,0\rangle,&\cr&
T_{0}|h,\overline{h},1\rangle\;=\;(h+\overline{h})|h,\overline{h},1\rangle
+|h,\overline{h},0\rangle\;=\;\mathcal{H}|\mathcal{H},\xi,1\rangle+|\mathcal{H},\xi,0\rangle,&\cr&
M_{0}|h,\overline{h},1\rangle\;=\;\frac{h-\overline{h}}{c}|h,\overline{h},1\rangle+
\frac{1}{c}|h,\overline{h},0\rangle\;=\;\xi|\mathcal{H},\xi,1\rangle.
\end{split}\eea

So we end up with Jordan scaling and diagonal rapidity. However
considering a complex $\kappa$ we observe that Jordan form for
rapidity arises as well\cite{Hosseiny:2011}. Since LCFT can be
realized in the bulk at the critical point of TMG, we now ask if we
can follow contraction and realize LGCA as well.

\section{Topologically Massive Gravity at the critical point}

Three dimensional gravity with or without cosmological constant term
has no propagating modes, but when one adds higher derivative terms
to the action nontrivial degrees of freedom arise. One of the best
known models in this regard is Topologically Massive Gravity
\cite{{Deser:1981wh},{Deser:1982vy}} in which the higher derivative
term is Chern-Simons term. In other words

\bea I=\frac{1}{16\pi G}\int d^{3}x\sqrt{-g}\bigg[
R+\frac{2}{l^{2}}+\frac{1}{\mu}\mathcal{L}_{CS}\bigg] \eea

where

\bea
\mathcal{L}_{CS}=\frac{1}{2}\epsilon^{\lambda\mu\nu}\Gamma_{\lambda\sigma}^{\alpha}
\big[\partial_{\mu}\Gamma_{\alpha\nu}^{\sigma}+\frac{2}{3}
\Gamma_{\mu\tau}^{\sigma}\Gamma_{\nu\alpha}^{\tau}\big] \eea

In general adding higher derivative term causes the instability due
to the presence of ghost-like modes. Nevertheless one can show
\cite{Li:2008dq} that TMG at the critical value for coupling of
Chern-Simons term to Einstein-Hilbert term is stable above the
$AdS_{3}$ vacuum with Brown-Henneaux condition. With this boundary
condition the left-moving degrees of freedom are pure gauge and
related symmetry charges are zero.

The equations of motion of this action is

\bea\label{e.o.m.T.M.G} R_{\mu\nu}-\frac{1}{2}R
g_{\mu\nu}+\frac{1}{l^{2}}g_{\mu\nu}+ \frac{1}{\mu}C_{\mu\nu}=0 \eea

where Cotton tensor is

\bea C_{\mu\nu}=\epsilon_{\mu}^{\alpha\beta}\nabla_{\alpha}(
R_{\beta\nu}-\frac{1}{4}R g_{\beta\nu}) \eea

TMG on an asymptotically locally $AdS_3$ geometry may have a dual
CFT with these central charges

\bea c_{R}\;=\;\frac{3l}{2G_{N}}(\frac{\mu l+1}{\mu
l}),\;\;\;\;\;\;\; c_{L}\;=\;\frac{3l}{2G_{N}}(\frac{\mu l-1}{\mu
l}). \eea

So one can take the non-relativistic limit from both sides of this
duality and check if this correspondence is still valid in this
limit. Recently it is shown \cite{Bagchi:2010mw} that in the
non-relativistic limit, the form of two-point correlation functions
in both sides are equal and have GCA structure away from the
critical point. However equation of motion (\ref{e.o.m.T.M.G}), at
the critical point has a non-trivial solution
\cite{Grumiller:2008qz} that do not obey Brown-Henneaux conditions
which may be interpreted as the left-moving excitations. The general
solution of linearized equation of motion at the critical point in
the finite neighborhood of the conformal boundary at $\rho=0$ has
the form of \footnote{Note that if $b_{(0)ij}=0$ this solution obeys
Brown-Henneaux boundary condition and $AdS$ vacuum is the case for
which $b_{(0)ij}=0$ and $g_{(0)ij}=\delta_{ij}$}.

\bea\label{a.ex1}
ds^2=\frac{d\rho^2}{4\rho^2}+\frac{1}{\rho}g_{ij}dx^idx^j, \eea in
which \bea\label{a.ex2} g_{ij}=b_{(0) ij}\log(\rho)+g_{(0)
ij}+\left(b_{(2) ij}\log(\rho)+ g_{(2) ij}\right)\rho+\cdots\ , \eea

When $b_{(0)ij}$ is turned on, this solution has not structure of
$AlAdS$ space-time. But with machinery of AdS/CFT
\cite{Maldacena:1997re}, it is shown in \cite{Skenderis:2009nt}that
TMG at the critical point might be  dual to logarithmic conformal
field theory treating $b_{(0)ij}$ perturbativly. Two point
correlation functions will be:

\bea\label{T.M.G.l.c.f}\begin{split}
&<T_{\overline{z}\;\overline{z}}(z,\overline{z})T_{\overline{z}\;\overline{z}}(0)>
\;=\;\frac{3l/2G_{N}}{\overline{z}^{4}},\;\;\;\;\;\;\;\;\;<t_{z,z}(z,\overline{z})T_{zz}(0)>
\;=\;\frac{-3l/2G_{N}}{z^{4}}&\cr&<t_{zz}(z,\overline{z})t_{zz}(0)>\;=\;\frac{1}{2G_{N}}
\frac{-3B_{m}-11+6\ln(m^{2}|z^{2}|)}{z^{4}}.
\end{split}\eea


Starting from the above equation and performing contraction we
obtain:

\bea\begin{split}
&<T_{\overline{z}\;\overline{z}}T_{\overline{z}\;\overline{z}}>\;=\;\frac{3
l}{2G_{N}}\frac{1}{(t-\frac{x}{c})^{4}} \;=\;\frac{3
l}{2G_{N}}t^{-4}+\frac{6l}{G_{N}}t^{-4}\; \frac{x}{ct}+...,&\cr&
<T_{zz}t_{zz}>\;=\;\frac{-3
l}{2G_{N}}\frac{1}{(t+\frac{x}{c})^{4}}\;=\;\frac{-3
l}{2G_{N}}t^{-4}+\frac{6l}{G_{N}} t^{-4}\; \frac{x}{ct}+...,&\cr&
<t_{zz}t_{zz}>\;=\;\frac{3l}{G_{N}} \log(m^{2}t^{2})\;t^{-4}(1-4\;
\frac{x}{ct}+...)-\frac{(3B_{m}+11)
l}{2G_{N}}t^{-4}(1-4\;\frac{x}{ct}+...)
\end{split}\eea

It can be easily seen that by no means one can redefine the
parameters to keep spatial dimension in the contraction limit. It is
because in (\ref{T.M.G.l.c.f}) factors of logarithmic term and
$z^{-4}$ term are coupled to each other. In other words at the first
sight it seems that since TMG at the critical point corresponds
to a special subclass of LCFT's, it's contraction fails to give an
interesting result. However in \cite{Kogan:2002mw} a different
approach to LCFT was introduced. In this approach LCFT is obtained
as vanishing limit of left central charge. In
\cite{Skenderis:2009nt} the authors utilized this insight to
evaluate TMG/LCFT correspondence. While taking the contraction limit
of the final result (\ref{T.M.G.l.c.f}) fails to keep spatial
dimension, taking both limits simultaneously and carefully one can
help spatial dimension survive.

\section{Contraction of TMG/LCFT}

\subsection{Contraction of CFT at $c_{L}\rightarrow 0$}

In the approach presented in \cite{Kogan:2002mw} it is supposed that
as $c_{L}$ goes to zero, beside energy-momentum tensor there is
another operator $X$ with conformal dimension of
$(2+\eta(c_{L}),\eta(c_{L}))$ which in the limit approaches to
$(2,0)$. The non-vanishing two point-functions are

\bea\begin{split}
<T_{\overline{z}\;\overline{z}}&T_{\overline{z}\;\overline{z}}>\;=\;\frac{c_{R}}{\overline{z}^{4}},
\;\;\;\;\;\;<T_{zz}T_{zz}>\;=\;\frac{c_{L}}{z^{4}},&\cr&<XX>\;=\;\frac{1}{c_{L}}\frac{\alpha(c_{L})}
{z^{4+2\eta(c_{L})}\overline{z}^{2\eta(c_{L})}}.
\end{split}\eea

Defining $t_{zz}=-\frac{1}{c_{L}}X-\frac{1}{c_{L}}T_{zz}$ and
letting $c_{L}\rightarrow 0$, one approach LCFT two-point functions
\bea\begin{split}
<T_{zz}(z)t_{zz}(0,0)>=&\frac{b}{2z^{4}},\;\;\;\;\;\;\;\;\;\;\;\;<t_{zz}(z,\overline{z})t_{zz}
(0,0)>=\frac{-b\log(|z^{2}|)}{z^{4}},&\cr&\;\;\;<T_{zz}(z)T_{zz}(0)>=0.
\end{split}\eea
Now we try to redefine energy-momentum tensors and try to
consider both limits simultaneously. To do so we let

\bea\begin{split}
&T_{1}\;=\;\frac{1}{\sqrt{c_{L}}}T_{\overline{z}\;\overline{z}}+\frac{\beta(c_{L})}{c_{L}}
X+\frac{\gamma(c_{L})}{c_{L}}T
,&\cr&T_{2}\;=\;\frac{1}{c}(\frac{1}{\sqrt{c_{L}}}T_{\overline{z}\;\overline{z}}
-\frac{\beta(c_{L})}{c_{L}}X-\frac{\gamma(c_{L})}{c_{L}}T) ,&\cr&
T_{3}\;=\;T_{zz}+\sqrt{c_{L}}\;T_{\overline{z}\;\overline{z}},
\end{split}\eea

and observe by taking limits of $c_{L}\rightarrow 0$ and $c\rightarrow
\infty$ simultaneously we obtain

\bea\begin{split}
&<T_{1}T_{1}>\;=\;\frac{1}{c_{L}}<T_{\overline{z}\;\overline{z}}T_{\overline{z}\;\overline{z}}>
+\frac{\beta^{2}}{c^{2}_{L}}<XX>+\frac{\gamma^{2}}{c^{2}_{L}}<T_{zz}T_{zz}>&\cr&=
\frac{c_{R}}{c_{L}}t^{-4}(1+4\frac{x}{ct}+...)+\frac{\beta^{2}\frac{\alpha
}{c^{2}_{L}}t^{-4\eta}}{c_{L}}(1-4\frac{x}{ct}+...)+\frac{\gamma^{2}}{c_{L}}
t^{-4}(1-4\frac{x}{ct}+...)&\cr&=t^{-4}\bigg(\frac{c_{R}+\beta^{2}
\frac{\alpha}{c^{2}_{L}}t^{-4\eta}+\gamma^{2}}{c_{L}}\bigg)+
4t^{-4}\bigg(\frac{c_{R}-\beta^{2}
\frac{\alpha}{c^{2}_{L}}t^{-4\eta}-\gamma^{2}}{c_{L}}\bigg)\frac{x}{ct}
\end{split}\eea

Note that for simplifying we have implied the functionality
of $\alpha,\beta, \gamma, \eta$ to $c_{L}$. Now requiring

\bea\label{c.fi.side} \lim_{c_{L}\rightarrow 0}\bigg(\beta^{2}
\frac{\alpha}{c^{2}_{L}}t^{-4\eta}+\gamma^{2}\bigg)\;=\;-c_{R} \eea

we obtain\footnote{Which $"\prime"$ indicate derivative respect to $c_{L}$.}

\bea\begin{split} <T_{1}T_{1}>\;&=\;t^{-4}\lim_{c_{L}\rightarrow
0}\bigg[(\gamma^{2})^{\prime}+(\beta^{2}
\frac{\alpha}{c^{2}_{L}})^{\prime}\bigg]
+4t^{-4}\lim_{c_{L}\rightarrow
0,c\rightarrow\infty}(\frac{2c_{R}}{c_{L}}\frac{1}{c})
\frac{x}{t}&\cr&+t^{-4}\lim_{c_{L}\rightarrow
0}\bigg[-2\eta^{\prime} \beta^{2}\frac{\alpha}{c^{2}_{L}}\bigg]\ln
t^{2}
\end{split}\eea

As well we observe

\bea\begin{split}
<T_{1}T_{2}>\;&=\;\frac{1}{c}\bigg(\frac{1}{c_{L}}<T_{\overline{z}\;\overline{z}}
T_{\overline{z}\;\overline{z}}>-\frac{\beta^{2}}{c^{2}_{L}}<XX>-\frac{\gamma^{2}}
{c^{2}_{L}}<T_{zz}T_{zz}>\bigg)&\cr&=t^{-4}\bigg(\frac{c_{R}-\beta^{2}
\frac{\alpha}{c^{2}_{L}}t^{-4\eta}-\gamma^{2}}{c_{L}}\bigg)\frac{1}{c}
+4t^{-4}\bigg(\frac{c_{R}+\beta^{2}\frac{\alpha}{c^{2}_{L}}t^{-4\eta}+
\gamma^{2}}{c_{L}}\bigg)\frac{1}{c^{2}}\frac{x}{t}
\end{split}\eea

which at the limits results in

\bea
<T_{1}T_{2}>\;=\;t^{-4}\lim_{c_{L}\rightarrow 0, c\rightarrow\infty}(\frac{2c_{R}}{c_{L}}\frac{1}{c})
\eea

For other two-point functions we obtain

\bea
<T_{1}T_{3}>\;=\;t^{-4}(\lim_{c_{L}\rightarrow 0}\gamma+c_{R}),\;\;\;\;\;\;<T_{2}T_{3}>\;=\;<T_{3}T_{3}>\;=\;0.
\eea

If we define

\bea\begin{split} &\lim_{c_{L}\rightarrow
0}\bigg[(\gamma^{2})^{\prime}+(\beta^{2}
\frac{\alpha}{c^{2}_{L}})^{\prime}\bigg]\;=\;C_{1},\;\;\;\;\;\;
\lim_{c_{L}\rightarrow 0,
c\rightarrow\infty}(\frac{2c_{R}}{c_{L}}\frac{1}{c})
\;=\;C_{2},&\cr&\lim_{c_{L}\rightarrow 0}\bigg[-2\eta^{\prime}
\beta^{2}\frac{\alpha}{c^{2}_{L}}\bigg]\;=\;C_{3},\;\;\;\;\;\;\;\;\;\;\;\;\;\;\;
\lim_{c_{L}\rightarrow 0}\gamma+c_{R}\;=\;C_{4}.
\end{split}\eea

The two-point correlation functions of non-relativistic limit will
be

 \bea\begin{split}
&<T_{1}T_{1}>\;=\;C_{1}t^{-4}+4 C_{2}t^{-4}\frac{x}{t}+C_{3} t^{-4}
\ln t^{2},&\cr&<T_{1}T_{2}>\;=\;C_{2}t^{-4},\;\;\;\;\;\;\;\;<T_{1}T_{3}>\;=\;C_{4}t^{-4},&\cr&
<T_{2}T_{2}>\;=\;<T_{2}T_{3}>\;=\;<T_{3}T_{3}>\;=\;0
\end{split}\eea

As it can be observed with coupling both limits we derived
a much larger class of LGCA in which beside the time,
spatial dependency appears in two-point functions. Now
we try to consider taking parallel limits in gravity
side to see if we can observe spatial dimension in
two-point functions.


\subsection{Topologically Massive Gravity at $\mu l\rightarrow 1$}

Note that with the above procedure we can determine the relative
behavior of two limits; one is critical limit $\mu l\rightarrow 1$
and the other is contraction limit $c\rightarrow \infty$. Let us
review some important points of section (7) of
\cite{Skenderis:2009nt}. To find two-point correlation functions
with the AdS/CFT machinery, one needs to calculate the second
expansion of action $I_{2}$ and linearized equations of motion
around the groundstate, AdS vacuum, simultaneously. Linearized
equations of motion give the most general asymptotic form of the
solution: \bea
h_{ij}\;=\;h_{(-2\lambda)ij}\rho^{-\lambda}+h_{(0)ij}+h_{(2)ij}\rho
+h_{(2-2\lambda)ij}\rho^{1-\lambda}+h_{(2+2\lambda)ij}\rho^{\lambda+1}+...
\eea with \bea\label{def.lam} \mu l\;=\;2\lambda+1. \eea

Equations of motion of TMG are third order in derivatives. This
means that one needs three boundary condition to determine the
solution. The first condition is regularity in $\rho=\infty$ limit.
So, we are left with two other conditions which are reflected in two
boundary sources $h_{(-2\lambda)ij},h_{(0)ij}$, for which we define
the corresponding CFT operators $T_{ij}$ and $X_{ij}$. First we need
to substitute this asymptotic form to second expansion of action and
determine diverging part of the action. Then after finding the
proper covariant counterterms to vanish diverging part, we need to
substitute again the asymptotic form in renormalized action
$I_{2,\lambda,tot}=I_{2}+I_{c.t}$. In this step one can determine
the one-point functions of two corresponding CFT operators with
functionally differentiation of reonormalized action
$I_{2,\lambda,tot}$ with respect to the two sources
$h_{(-2\lambda)ij},h_{(0)ij}$.

\bea
<X_{ij}>\;=\;\frac{-4\pi}{\sqrt{-g_{AdS}}}\frac{\delta I_{2,\lambda,tot}}
{\delta h^{ij}_{(-2\lambda)}},\;\;\;\;\;\;\;\;\;<T_{ij}>\;=\;\frac{4\pi}{\sqrt{-g_{AdS}}}
\frac{\delta I_{2,\lambda,tot}}{\delta h^{ij}_{(0)}}.
\eea

To find the two-point functions one needs to functionally
differentiate one-point functions with respect to the sources.
However in one-point functions there exist some terms that their
forms are not fully determined by asymptotic analysis. In this step
of holographic renormalization one needs to solve the linearized
equations of motion exactly to fully determine the dependency of
these terms to the sources. After following this procedure, we end
up with these two-point functions\footnote{One first uses lightcone
coordinates $v,u=t\pm x$ and replaces $v\rightarrow z$,$u\rightarrow
\overline{z}$, the complex boundary coordinates.}.
\bea\label{T.M.G.c.f}\begin{split}
&<T_{\overline{z}\;\overline{z}}(z,\overline{z})T_{\overline{z}\;\overline{z}}(0)>\;=\;
\frac{3l}{2G_{N}}\frac{\lambda+1}{2\lambda+1}\frac{1}{\overline{z}^{4}},
&\cr&<T_{zz}(z,\overline{z})T_{zz}(0)>\;=\;\frac{3l}{2G_{N}}\frac{\lambda}{2\lambda+1}\frac{1}{z^{4}},&\cr&
<X_{zz}(z,\overline{z})X_{zz}(0)>\;=\;-\frac{l}{2G_{N}}\frac{\lambda(\lambda+1)(2\lambda+3)}{2\lambda+1}
\frac{1}{z^{2\lambda+4}\overline{z}^{2\lambda}}.
\end{split}\eea

Defining

\bea\label{T.M.G.c}
T_{1}\;=\;\frac{1}{\sqrt{\lambda}}T_{\overline{z}\;\overline{z}}-\frac{1}{\lambda}X,
\;\;\;\;\;\;T_{2}\;=\;\frac{1}{c}(\frac{1}{\sqrt{\lambda}}T_{\overline{z}\;\overline{z}}+\frac{1}{\lambda}X),
\;\;\;\;\;\;T_{3}\;=\;T_{zz}+\sqrt{\lambda}\;T_{\overline{z}\;\overline{z}}.
\eea

and looking over (\ref{T.M.G.c.f})and (\ref{def.lam}), we obtain

\bea\begin{split}
<T_{1}T_{1}>&\;=\;\frac{\mu l+1}{2\mu l(\mu l-1)}\bigg(\frac{3l}{G_{N}}\frac{1}{\overline{z}^{4}}
-\frac{l}{G_{N}}(\mu l+2)\frac{1}{z^{(\mu l-1)+4}\overline{z}^{\mu l-1}}\bigg)&\cr&=\frac{\mu l+1}
{2\mu l(\mu l-1)}t^{-4}\bigg(\frac{3l}{G_{N}}[1+4\frac{x}{ct}+...]-\frac{l}{G_{N}}(\mu l+2)
[1-4\frac{x}{ct}+...]t^{-2(\mu l-1)}\bigg)&\cr&=\frac{\mu l+1}{2\mu l}t^{-4}\bigg(\frac{l}{G_{N}}
\frac{3-(\mu l+2)t^{-2(\mu l-1)}}{\mu l-1}+4\frac{l}{G_{N}}\frac{3+(\mu l+2)t^{-2(\mu l-1)}}{\mu l-1}
\frac{x}{ct}\bigg)
\end{split}\eea

considering

\bea\begin{split}
&\lim_{\mu l\rightarrow 1}\frac{3-(\mu l+2)t^{-2(\mu l-1)}}{\mu l-1}\;=\;3\ln t^{2}-1,&\cr&
\lim_{\mu l\rightarrow 1,c\rightarrow \infty}\frac{3+(\mu l+2)t^{-2(\mu l-1)}}{\mu l-1}\frac{1}{c}\;=\;6.
\end{split}\eea

we end up

\bea
<T_{1}T_{1}>\;=\;t^{-4}\bigg(-\frac{l}{G_{N}}+4\frac{6l}{G_{N}}\frac{x}{t}+\frac{3l}{G_{N}}\ln
t^{2}\bigg). \eea

As well we observe \bea\begin{split} <T_{1}T_{2}>&\;=\;\frac{\mu
l+1}{2\mu l(\mu
l-1)c}t^{-4}\bigg(\frac{3l}{G_{N}}[1+4\frac{x}{ct}+...]
+\frac{l}{G_{N}}(\mu l+2)[1-4\frac{x}{ct}+...]t^{-2(\mu
l-1)}\bigg)&\cr&=\frac{\mu l+1}{2\mu l}t^{-4}\bigg(\frac{l}{G_{N}}
\frac{3+(\mu l+2)t^{-2(\mu l-1)}}{\mu
l-1}\frac{1}{c}+4\frac{l}{G_{N}}\frac{3-(\mu l+2)t^{-2(\mu
l-1)}}{\mu l-1} \frac{x}{c^{2}t}\bigg)&\cr&=t^{-4}(\frac{6l}{G_{N}})
\end{split}\eea

and

\bea
<T_{1}T_{3}>\;=\;\frac{3l}{G_{N}},\;\;\;\;\;\;\;\;\;\;
<T_{2}T_{2}>\;=\;<T_{2}T_{3}>\;=\;<T_{3}T_{3}>\;=\;0.
\eea

Comparing with the forms of two-point functions of LGCA introduced
in previous section

\bea\begin{split}
&<T_{1}T_{1}>\;=\;C_{1}t^{-4}+4 C_{2}t^{-4}\frac{x}{t}+C_{3} t^{-4}
\ln t^{2},&\cr&<T_{1}T_{2}>\;=\;C_{2}t^{-4},\;\;\;\;\;\;\;\;<T_{1}T_{3}>\;=\;C_{4}t^{-4},&\cr&
<T_{2}T_{2}>\;=\;<T_{2}T_{3}>\;=\;<T_{3}T_{3}>\;=\;0
\end{split}\eea

We conclude TMG at the critical point can be dual to LGCA with the
following central charges

\bea
C_{1}\;=\;-\frac{l}{G_{N}},\;\;\;\;\;\;C_{2}\;=\;\frac{6l}{G_{N}},
\;\;\;\;\;\;C_{3}\;=\;\frac{3l}{G_{N}} ,\;\;\;\;\;\;C_{4}\;=\;\frac{3l}{G_{N}}.
\eea

\section{Conclusions and Outlook}

In this paper we explored some aspects of TMG/LGCA holography. We
observe that for this duality to exist at least at the level of
two-point correlation functions, we should carefully take both
limits. One limit is the contraction limit in which the speed of
light $c$ goes to infinite and the other is the "chiral" limit $\mu
l\rightarrow 1$ that corresponds to vanishing left central charge.
Taking both limits separately, contracted two-point correlation
functions depends only on time. However considering both limits
simultaneously; $\mu l-1$ coupled to $\frac{1}{c}$; we observe
existence of spatial dimension in final results. It means that when
contracting a model one needs to be careful to consider all possible
limits to span the space of all variable of the contracted model more thoroughly.\\
We should notify that when an algebra is contracted there is no necessity
that contracting its representation as well yields thorough 
representation of the contracted model. We can utilize this apparatus 
if we have a good knowledge about the contracted model. For
$CFT_{2}/GCA_{2}$ since previously many aspects of its representation
and as well its two point functions \cite{{Henkel:2006},{Bagchi:2009ca}} were known, this
approach had something strong to say. Beside two-point functions
many more reasons were presented in \cite{Bagchi:2009pe} to
support this approach. Otherwise, contraction approach may span
only a subclass of representations of the contracted algebra and just give
some insights about the contracted model. So we should mention that though our final contraction
yields spatial term for two-point functions but a 
concrete statement is subject to more thorough investigation.\\
Another three dimensional gravity with higher derivative terms has
been introduced in \cite{Bergshoeff:2009hq}.
 The corresponding action is given by
\bea\label{action} S&=&\frac{1}{16\pi G_{N}}\int_{R}
d^3x\sqrt{-G}\left[R-2 \lambda\right]\
-\frac{1}{m^{2}}\frac{1}{16\pi G_{N}}\int_{R}
d^3x\sqrt{-G}\left[R^{\mu \nu}R_{\mu \nu}-\frac{3}{8}R^{2}\right].
\eea This model is known as New Massive Gravity (NMG). It is
believed that NMG model on an asymptotically locally $AdS_3$
geometry may have a dual CFT whose central charges are given by
\cite{{Bergshoeff:2009aq},{Liu:2009kc}}( see also
\cite{{Bergshoeff:2009tb},{Liu:2009bk}}) \bea
c_L=c_R=\frac{3l}{2G_{N}}\left(1-\frac{1}{2m^2l^2}\right). \eea At
the critical value, $m^2l^2=\frac{1}{2}$, where the central charges
are zero it has been shown \cite{AyonBeato:2009yq} that the model
admits a new vacuum solution which is not asymptotically locally
$AdS_3$. Near the boundary, form of this solution is similar to
TMG's solution in critical point
(\ref{a.ex1}),(\ref{a.ex2}),however, relations between the
coefficients are different due to the dynamical differences of two
theories. It is shown that NMG at the critical point
\cite{{Grumiller:2009sn},{Alishahiha:2010bw}} may be dual to LCFT.
In contrast to TMG, this model has parity symmetry, so the left and
right central charges are equal. As a result this model may not be
applicable for studying NMG/GCA holography. However the thorough
examination is left : If we can by the contraction procedure
followed in this paper introduce proper NMG/LGCA holography or not.

\section*{Acknowledgments}

We would like to thank M. Alishahiha, R. Fareghbal, A. Mosaffa , S.
Rouhani and A. Vahedi for useful discussions. This work has been
supported by Iran National Science Foundation (INSF).



\end{document}